\documentclass[11pt]{article}
\usepackage{graphicx}
\usepackage{amsmath}
\usepackage{amssymb}
\usepackage{amsxtra}
\def\({\left(}
\def\){\right)}

\def\beq{\begin{equation}}
\def\eeq{\end{equation}}
\def\bea{\begin{eqnarray}}
\def\eea{\end{eqnarray}}
\title{Recent advances of Beyond the Standard model cosmology}
\author{Maxim Khlopov\\
Virtual Institute of Astroparticle physics, 75018 Paris, France\\ 
Center for Cosmoparticle physics ``Cosmion''\\ National Research Nuclear University MEPhI, 115409 Moscow, Russia\\
Research Institute of Physics, Sousthern Federal University,\\ Stachki 194, Rostov on Don 344090, Russia,\\ 
e-mail khlopov@apc.univ-paris7.fr}

\begin{document}
\maketitle

\begin{abstract}
BSM physics, underlying the now standard inflationary cosmology with baryosynthesis and dark matter/energy, inevitably leads to features, going beyond this standard cosmological paradigm. The signatures of these features can be considered in the hints for the existence of antihelium component of cosmic rays and massive Primordial Black Holes (PBH), in the discovery of stochastic gravitational wave background (SGWB) and evidence for early galaxy formation as well as in the dark atom solution for the puzzles of direct dark mater searches. We discuss open problems of physics of dark atoms, which can explain positive result of DAMA/NaI and DAMA/LIBRA direct searches for dark matter, of the origin and evolution of antimatter domains in baryon asymmetrical Universe, of the Axion Like Particle (ALP) physics, which may explain the origin of SGWB, discovered by Pulsar Timing Array (PTA) facilities and early galaxy formation, favored by James Webb Space Telescope. 
\end{abstract}


\section{Introduction}\label{s:intro}
The now Standard cosmological paradigm involves inflation, baryosynthesis and dark matter/energy \cite{PPNP,Lindebook,Kolbbook,Rubakovbook1,Rubakovbook2,book,newBook,4,DMRev}, which imply physics Beyond the Standard Model (BSM) of fundamental interactions. Model dependent cosmological consequences of this physics give its cosmological messengers \cite{Bled21,Protvino,ecu}, which challenge the concordance of the standard $\Lambda$CDM cosmology and can remove the conspiracy of its BSM features \cite{ijmpd19}. Positive hints to these features need more detailed analysis of physical basis and observable features of BSM messengers.  Here we discuss observational signatures of BSM cosmology presented at the XXVI Bled Workshop "What comes beyond the Standard models?" with special emphasis on open problems of their analysis.

Confirmation of these signatures would strongly reduce the set of models and parameters of the physical basis of the modern cosmology. Therefore, mechanisms of inflation and baryosynthesis as well as dark matter candidates proposed in any approach to the unified description of Nature \cite{Norma,Norma2} should inevitably include BSM cosmological signatures, which find observational support. 

We consider open questions in dark atom messengers of new physics in the direct searches of dark matter (Section \ref{da}), in the footprints of Axion Like Particle (ALP) physics in stochastic gravitational wave background (SGWB) and early galaxy formation, favored by James Webb Space Telescope (JWST) and in the primordial objects of antimatter, which can be sources of antihelium component of cosmic rays (Section \ref{pbh}). We discuss is the conclusive Section \ref{cpp} there signatures and their significance in the context of cosmoparticle physics of BSM physics and cosmology.
\section{Open problems of dark matter physics}
\subsection{Dark atom signature in direct dark matter searches}\label{da}
The increasingly high statistics of positive result of underground direct dark matter search in DAMA/NaI and DAMA/LIBRA experiments \cite{rita} challenges its  Weakly Interacting Massive Particles (WIMP) interpretation with the account for negative results of direct WIMP searches by other groups (see \cite{PPNP} for review and references). Though the difference in experimental strategy may leave some room for WIMP interpretation of this positive result \cite{rita}, its non-WIMP interpretation deserves serious attention and can make this result an experimental evidence for dark atom nature of dark matter \cite{PPNP,ecu,I,kuksa}.

The dark atom hypothesis assumes existence of stable particles with negative even electric charge $-2n$, which bind with $n$ primordial helium-4 nuclei in neutral nuclear interacting atom like states. 

The motivation for the existence of such multiple charged stable states can come from the composite nature of Higgs boson. Indeed, the lack of positive evidence for supersymmetric (SUSY) particles at the LHC can indicate very high energy SUSY scale \cite{ketovSym}. It makes hardly possible to use SUSY for solution of the problem of the divergence of Higgs boson mass and the origin of the energy scale of the electroweak symmetry breaking and implies a non-SUSY solution of this problem. Such a non-SUSY solution can be provided by composite Higgs boson and if Higgs constituents are charged, their composite multiple charged states can naturally provide  $-2n$ charged constituents of dark atoms, as it takes place in Walking Technicolor model (WTC). Nontrivial electroweak charges of techniparticles provide due to electroweak sphaleron transitions balance between their charge asymmetry (excess of $-2n$ charged particles over their antiparticles) and baryon asymmetry, thus predicting relationship between dark matter density in the form of dark atoms and baryon density. The model involves only one parameter of BSM physics - mass of the stable $-2n$ charged particles and reproduction of the observed  dark matter density makes possible to determine this parameter \cite{arnab,sopin}. It is shown in \cite{arnab,sopin} that the excess of $-2n$ charged techniparticles and -2 charged ($\bar U \bar U \bar U$) clusters of stable antiquarks $\bar U$ with the charge $-2/3$ of new stable generation like the 5th generation in the approach \cite{Norma2}, can explain the dark matter. The excess generated by such balance depends on the mass of multiple charged particles. It can put upper limit on the mass of stable multiple charged particles, at which dark atoms can explain the observed density of dark matter \cite{sopin}.

Dark atom structure depends on the value of $-2n$ charge. Double charged particles ($n=1$) form with primordial helium nucleus  Bohr-like $O$He atom, in which radius of helium Bohr orbit is nearly equal to the sizes of this nucleus. At $n>1$ particle with the charge $-2n$ forms with $n$ helium nuclei Thomson-like $X$He atom, in which $-2n$ charged lepton is situated within an $n$-$\alpha$-particle nucleus. In the both cases dark atoms structure strongly differs from the usual atoms. Instead of small nuclear interacting core and large electroweakly interacting shell they have a heavy lepton or lepton-like core and nuclear interacting helium shell. The latter determines their interaction with baryonic matter.

Dark atom hypothesis can qualitatively explain negative results of direct WIMP searches. Owing to their nuclear interaction they are slowed down in the terrestrial matter. It leads to negligible nuclear recoil in the underground detectors \cite{PPNP,ecu,I}. 

At each level of terrestrial matter dark atom concentration is determined by the balance between the incoming cosmic flux of dark atoms and their diffusion towards the center of Earth. At the 1 km depth this equilibrium concentration is adjusted to the incoming cosmic flux at the timescale of less than 1 hour. It leads to annual modulation of this concentration within the underground detector.  If dark atoms can form low energy (few keV) bound states with nuclei of detector, the energy release in such binding should possess annual modulation. It can explain the signal, detected in DAMA/NaI and DAMA/LIBRA experiments.

\begin{figure}
    \begin{center}
        \includegraphics[scale=0.6]{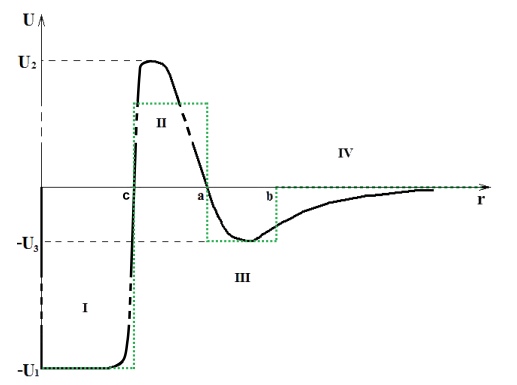}
        \caption{Potential barrier in dark atom- nucleus interaction can lead to a shallow potential well in which a low energy bound state can exist }
        \label{q1}
    \end{center}
\end{figure}

The crucial point in this explanation is the existence of a potential barrier and a shallow well in the interaction of a nucleus and dark atom. Qualitatively the origin of this barrier can be related with the specific interplay of Coulomb repulsion and nuclear attraction between the nucleus of detector and helium shell of dark atom. However in the lack of usual approximations of atomic physics (small ratio of nuclear to atomic sizes and electroweak coupling of electronic shell) quantitative description of dark atom interaction implies development of special numerical methods. In this direction numerical methods of continuous extension of a classical three body problem to realistic quantum-mechanical description were developed \cite{timur} both for Bohr-like and Thomson-like dark atoms interaction with nuclei \cite{timur2}. Presently numerical solutions for Schrodinger equation for dark-atom -nucleus quantum system are under way to make possible interpretation of the results \cite{rita} in terms of signature of dark atoms.
\subsection{Axion-like particle models of dark matter}
Axion-Like Particle (ALP) models can be reduced to a simple model of a complex field $\Psi = \psi \exp{i \theta}$ with broken global $U(1)$ symmetry \cite{PPNP,Protvino}. The potential
$$V=V_0+\delta V$$
 contains the term 
 \beq V_0= \frac{\lambda}{2}(\Psi^{\ast} \Psi -f^2)^2 \label{vo} \eeq
 that leads to spontaneous breaking of the $U(1)$ symmetry with continuous degeneracy of the asymmetric ground state
\beq \Psi_{vac}= f \exp (i \theta) \label{cav} \eeq
 and the term 
 \beq \delta V (\theta)= \Lambda^4 (1-\cos\theta) \label{disc} \eeq\
 with $\Lambda \ll f$ that leads to manifest breaking of the residual symmetry, leading to a discrete set of degenerated ground states, corresponding to 
$$\theta_{vac} = 0, 2\pi, 4\pi, ...$$ 
In the result of the second step of symmetry breaking an ALP field $\phi = f \theta$ is generated with the mass \beq m_{\phi} = \Lambda^2/f. \label{pngm} \eeq
 
 The term (\ref{disc}) can be present in the theory initially. Then coherent oscillations of the ALP field start, when Hubble parameter $H= m_{\phi} = \Lambda^2/f$. This term can be generated by instanton transitions, as it is the case in the axion models. Then these oscillations are switched on, when this term is generated at $T \sim \Lambda$ (i.e. at $H=\Lambda^2/m_{Pl}$). In spite of a very small mass of ALP, they are created initially nonrelativistic and thus behave like Cold Dark matter in the process of growth of small density fluctuations at the matter dominated stage. However, as we can see in the next section \ref{pbh}, ALP field evolution may be accompanied by creation of strong primordial inhomogeneities of different kind.

\section{Strong primordial inhomogeneities from ALP physics}\label{pbh}
\begin{figure}
    \begin{center}
        \includegraphics[scale=0.6]{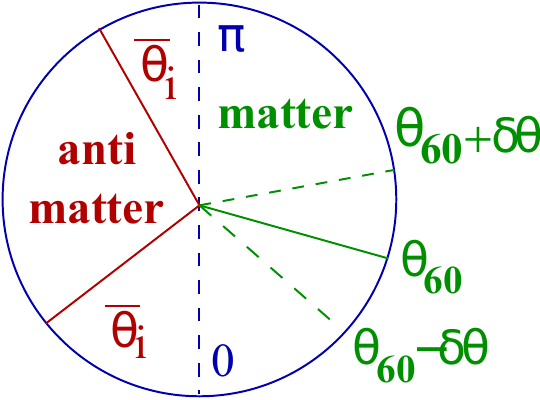}
        \caption{Phase fluctuations at inflationary stage can cross $\pi$, leading to formation of closed domain walls. In the model of spontaneous baryosynthesis, based on ALP physics phase fluctuations, crossing $\pi$ and 0, result in antibaryon domains in baryon asymmetric Universe \cite{PPNP} }
        \label{a}
    \end{center}
\end{figure}

If the first phase transition takes place during inflation with Hubble parameter $H_{\rm infl}$, quantum fluctuations within the scalar field $\Psi$ led to variations in phase between disconnected regions, quantified by $\delta\theta = H_{\rm infl}/(2\pi f)$. In regions where inflationary dynamics have resulted in phase values greater than $\pi$, the field will oscillate around the minimum at $\theta_{vac} = 2\pi$. Conversely, in the surrounding space where phase values are less than $\pi$, the oscillations will favor the minimum at $\theta_{vac} = 0$. This led to the formation of closed domain walls with $\theta = \pi$, which are closed surfaces in space. The phase variation process can be compared to a one-dimensional Brownian motion, where the field's phase undergoes a random walk with step lengths of approximately $\delta\theta$ per Hubble time. If the field $\Psi$ possess interactions with quarks and leptons, in which baryon and lepton numbers are not conserved, decay of the field $f\theta$ in its motion to ground state should generate baryon asymmetry, if $\theta < \pi$, or excess of antibaryons, if $\theta > \pi$ (see Fig. \ref{a}).

\subsection{ALP signatures in PBH, SGWB and JWST}
Formation of closed domain walls can lead to formation of primordial black holes (PBH) \cite{ketovSym,PBHrev}. Their mass $M$ is determined by the two fundamental scales of the ALP physics, $f$ and $\Lambda$, as given by 
\beq M_{min} = f(m_{Pl}/\Lambda)^2 \le M \le M_{max} = f (m_{Pl}/f)^2 (m_{Pl}/\Lambda)^2. \label{pbhalp}\eeq
Here $M_{min}$ is determined that the gravitational radius of wall $r_g = 2 M/m_{Pl}^2$ exceeds the width of wall $d \sim f/\Lambda^2$, while the maxiomal mass is determined by the condition that the wall can enter horizon before it starts to dominate in the energy density within horizon.  This mechanisms of PBH formation can provide formation of PBHs with stellar mass, and even larger than stellar up to the seeds for Active Galactic nuclei (AGN) \cite{AGN,RubinCluster,dolgovPBH}. LIGO/VIRGO detected gravitational wave signal from coalescence of black holes with masses  ($M > 50 M_{\odot}$), which exceed the limit of pair instability, and put forward the question on their primordial origin \cite{LV150,LV150apl}. 

\begin{figure}
    \begin{center}
        \includegraphics[scale=0.3]{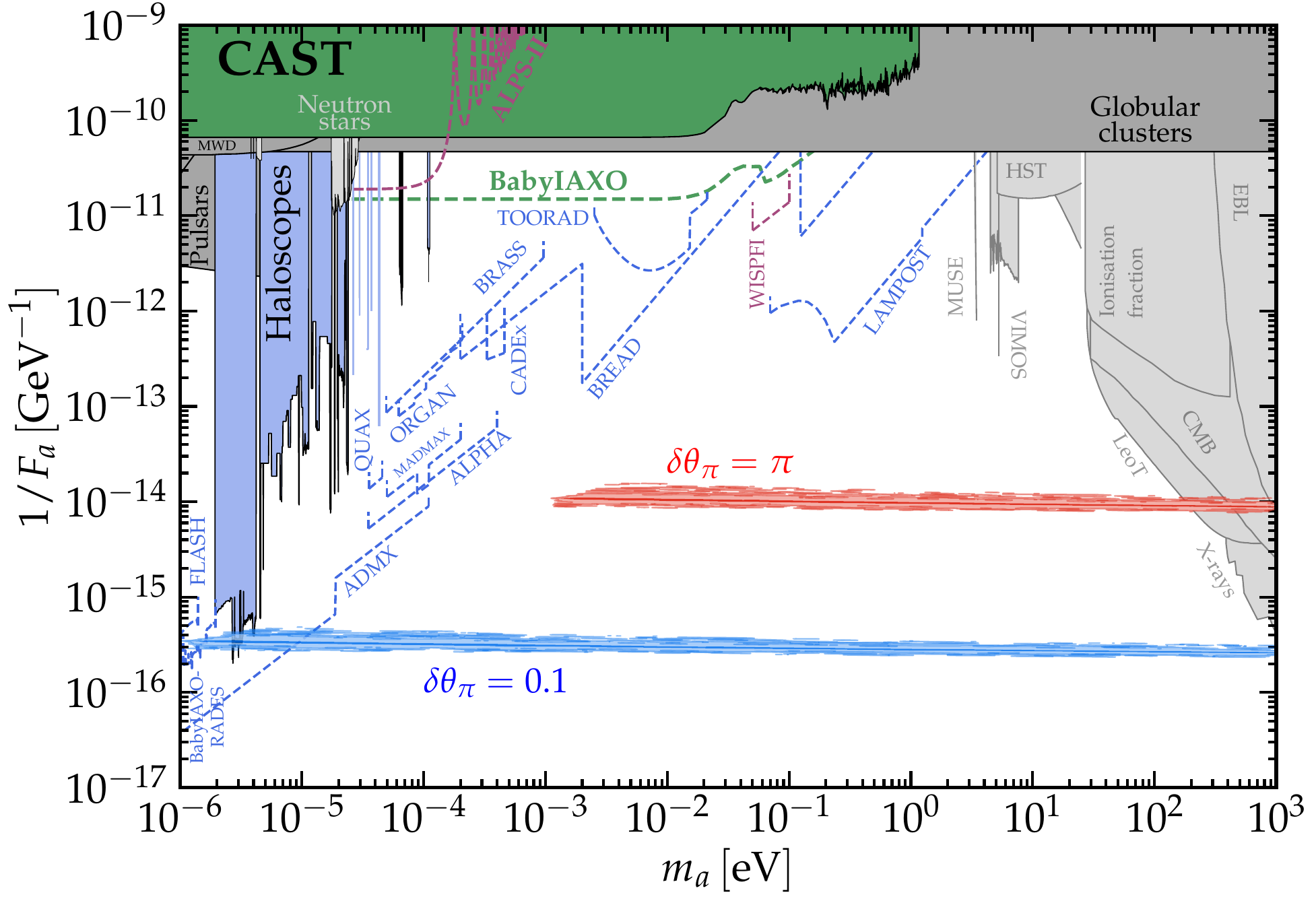}
        \caption{The possibility to explain SGWB and JWST data in ALP model (taken from \cite{shu,shu2}).}
        \label{q}
    \end{center}
\end{figure}

Recent discovery of Stochastic Gravitational Wave Background (SGWB) by Pulsar Timing Arrays (PTA) \cite{nano} can be another evidence of ALP physics. Collapse of large closed domain walls with mass $M>M_{max}$ leads to separation of the region from the surrounding Universe with formation of a wormhole and then baby Universe. This process is accompanied by gravitational wave background radiation, which can reproduce the PTA data. Simultaneously, the values of phase at the stages of inflation, preceding the stage of crossing $\pi$, at which the contour of the future domain wall is created, approach to $\pi$ and result in the energy density much larger, than the average one for the ALP field. It leads to the high ALP density in the regions, surrounding the wall and even, if the wall disappears in the baby Universe the ALP density in the surrounding region is much higher than the average in the Universe. It strongly facilitates galaxy formation at the redshifts $z>10$, indicated by the data of JWST, in the regions surrounding large closed domain walls. In that way ALP physics can simultaneously explain the PTA and JWST data \cite{shu,shu2}. The possibilities to probe such ALP physics are illustrated on Fig. \ref{q}. The open question in this scenario is the evolution of the regions of the enhanced ALP density, and, in particular, whether black hole formation is possible in their central part, or the enhanced ALP density itself plays the role of AGN seeds of early galaxies. 

\subsection{Antimatter domains in baryon asymmetric Universe}
Baryon asymmetry of the Universe, reflecting the absence of the comparable with baryonic matter amount of macroscopic antimatter in the observed Universe. Its origin is ascribed to the mechanism of baryosynthesis, in which baryon excess is created in very early Universe. Inhomogeneous baryosynthesis can lead in the extreme case to the change of sign of this excess, giving rise to antimatter domains, produced in the same process, in which the baryonic matter was created \cite{CKSZ,DolgovAM,Dolgov2,Dolgov3,KRS2,AMS}. Surrounded by matter, antimatter domains should be sufficiently large to survive to the present time and to give rise to antimatter objects in the Galaxy. It implies also effect of inflation in addition to nonhomogeneous baryosynthesis. Such combination of inflation and nonhomogeneous baryosynthesis can take place in spontaneous baryosynthesis at the specific choice of ALP parameters. 

This choice should determine the initial properties of antimatter domains and the forms of macroscopic antimatter, which can be formed in them.  The estimated minimal mass of surviving domain together with upper limit on possible amount of antimatter in our Galaxy from the observed gamma radiation gave the interval of mass $10^3 M_{\odot} \le M \le 10^5 M_{\odot}$, which is typical for globular clusters.  Symmetry of electromagnetic and nuclear interactions of matter and antimatter, space distribution of globular clusters in galactic halo with low gas density seemed to favor the hypothesis of antimatter globular cluster in our Galaxy\cite{AMS,GC}. The antibaryon density may be much higher, than the baryonic density and then specific ultra-dense antibaryon stars can be formed \cite{Blinnikov}, 

The predicted fraction of antihelium nuclei in cosmic rays from astrophysical sources is far below the sensitivity of AMS02 experiment \cite{poulin} . It makes the unpublished suspected antihelium-4 event, registered in 2017, a strong signature of macroscopic antimatter in our Galaxy. The unprecedented sensitivity of AMS02 experiment to the cosmic ray fluxes makes this collaboration especially responsible for presentation of its results, which cannot be tested by any other experimental group. That is why the collaboration continues to gain more and more statistics and check all the possible background interpretation before the discovery of cosmic antihelium-4 is announced.

With the hope that such an announcement will be made, studies of possible forms of macroscopic antimatter objects in our Galaxy are challenging. 

Such analysis should involve evolution of antibaryon domains in baryon asymmetrical universe \cite{orch,orch2} in the context of models of nonhomogeneous baryosynthesis. 

The earlier idea to use the observed properties of the M4 globular cluster as possible prototype of antimatter object \cite{nastya} should be strongly corrected, since chemical evolution of isolated antimatter domain should strongly differ from such evolution of the ordinary matter. One can expect that primordial nucleosynthesis should lead to production of primordial antihelium at rather wide range of antibaryon density in the domain, but the products of stellar anti-nucleosynthesis cannot come to the domain from other parts of the Galaxy, while heavy elements produced by antimatter stars within it leave domain and annihilate with the matter in the Galaxy. It makes hardly possible enrichment of antimatter object by antinuclei heavier than antihelium-4, while  the observed metallicity in all the galactic globular clusters is close to the Solar one, favoring the income of products of stellar nucleosynthesis from other parts of the Galaxy.

Primordial metallicity can appear in domains with high antibaryon density, in which antinuclei much heavier than helium-4 can be produced. In the context of nonhomogeneous nucleosynthesis based on ALP physics such high density antibaryon domains can appear after crossing $\pi$ and should have massive domain walls at their border. Collapse of such walls with the mass $M > M_{max} = f (m_{Pl}/f)^2 (m_{Pl}/\Lambda)^2$ puts such domains within baby Universes, so that only domains surrounded by walls with $M < M_{max}$ can be observable and there is an open question, whether such domains are sufficiently large to survive in the matter surrounding.  .

In any case to confront AMS02 searches the composition and spectrum of cosmic antinuclei from antimatter objects in our Galaxy should be predicted with the account for propagation in galactic magnetic fields of antinuclei from local source in the Galaxy \cite{nastya2}  
\section{Conclusions}\label{cpp}
The increasing hints to new physics phenomena in DAMA experiments, LIGO-VIRGO-KAGRA, PTA and JWST data, possible existence of antihelium component of cosmic rays can indicate not only effects of BSM physics, but also lead to the BSM cosmology, involving such deviations from the standard cosmological model as Warmer-than-Cold dark matter scenario of nuclear interacting dark atoms of dark matter, or primordial strong nonhomogeneities of energy and/or baryon density, giving rise to new scenarios of galaxy formation and evolution.  We have outlined here the open questions in the proposed BSM models and scenarios, which can explain these deviations and deserve special studies and discussion at future Bled Workshops.

In the context of cosmoparticle physics, studying fundamental relationship of macro- and micro- worlds in the cross-disciplinary studies of of its physical, astrophysical and cosmological signatures, confirmation of these cosmological messengers of new physics would provide a sensitive probe for BSM cosmology based on the proper choice of parameters of proper class of BSM models, since only such models, which predict these deviations from the standard cosmological paradigm can pretend to be realistic in this case. 
\section*{Acknowledgements}
The work by M.K. was performed with the financial support provided by the Russian Ministry of Science and Higher Education, project “Fundamental and applied research of cosmic rays”, No. FSWU-2023-0068.

%% The bibliography section

\end{document}